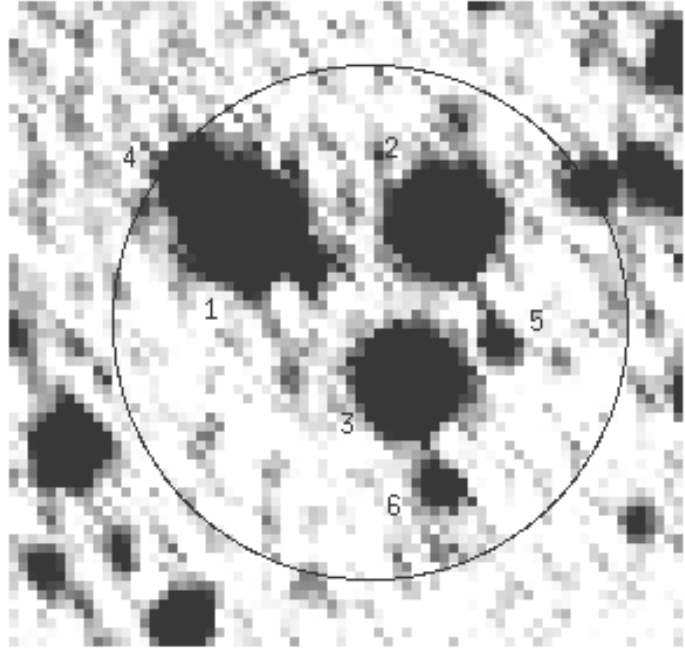

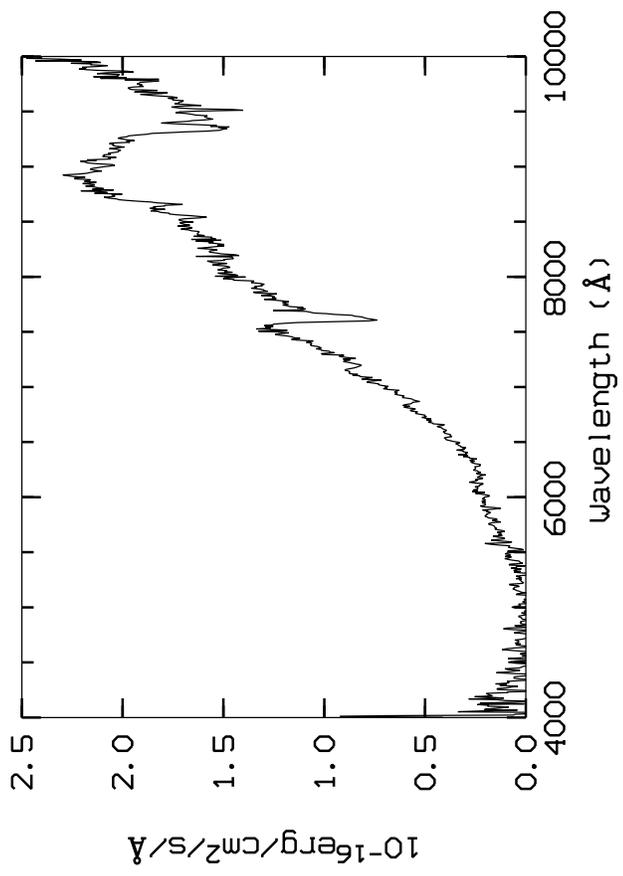

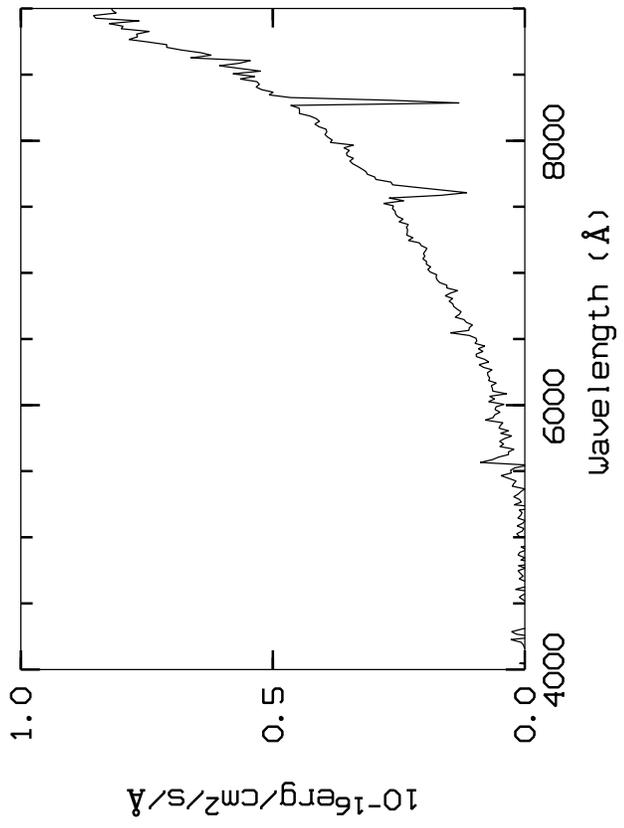

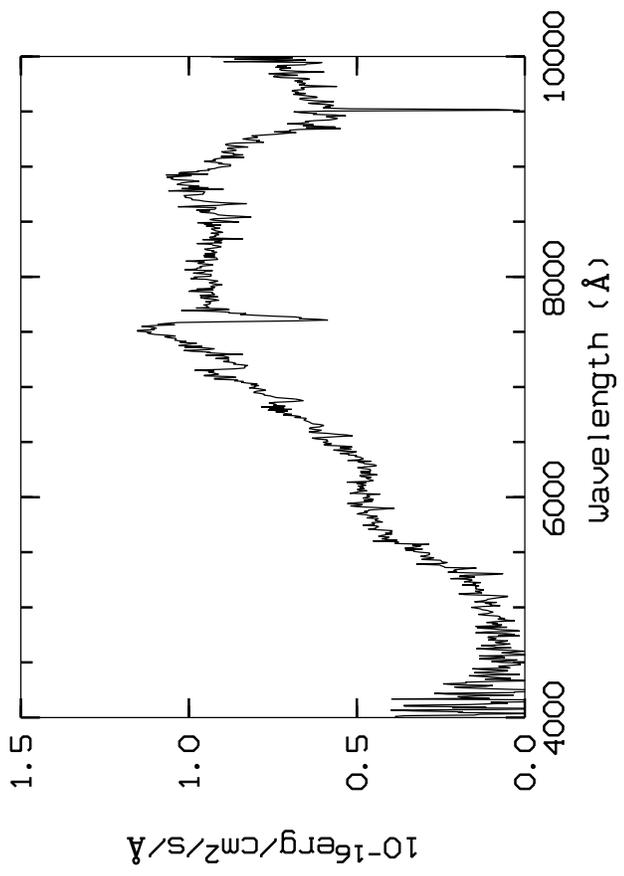

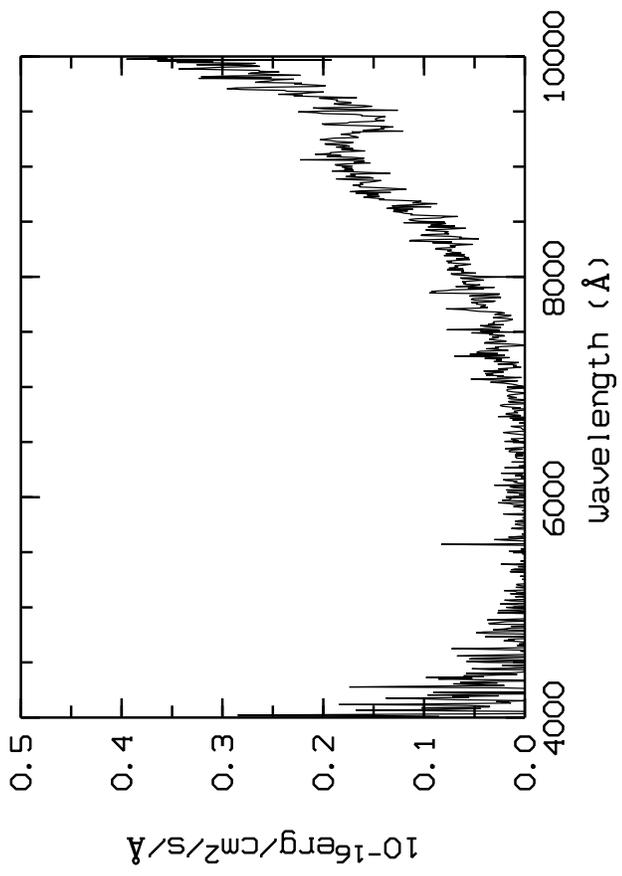

# THE OPTICAL CONTENT OF THE ERROR BOX OF GRS 1915+105

M. Boër*
Centre d'Etude Spatiale des Rayonnements (CNRS/UPS),
BP 4346, 31029 Toulouse Cedex, France

J. Greiner
Max Planck Institut für Extrarrestrische Physik
D 85740 Garching, FRG

C. Motch
Observatoire de Strasbourg
11, rue de l'Université, 67000 Strasbourg, France

**Abstract :**
   We report on deep photometry and spectroscopy of the objects detected in the ROSAT HRI X-ray error box of the hard X-ray transient, GRS 1915+105. The probable counterpart of this source has been detected, with a magnitude of 23.4 in the I band. The optical absorption $A_V = 28$ derived from the X-ray data implies an absolute magnitude in the I band, $M_I \approx -6.7$. We present the data for all the objects present in the error box, together with the spectra of the three brightest. Possible spectral types, as well as estimates of the hydrogen column density are derived.

---

# 1. Introduction

The GRS 1915+105 source was discovered in August 1992 by the WATCH instrument aboard the GRANAT spacecraft (Castro - Tirado et al., 1992). Until the end of 1994, this object has shown variable activity, with an average of 260 mCrab, and a maximum around 600mCrab (Castro-Tirado et al., 1994; Greiner et al., 1994; Harmon et al., 1994). A softening of the spectrum, from a spectral index of 2 to 2.8 was also observed by the BATSE instrument on-board the Compton GRO observatory (Greiner et al., 1994). This source was observed by the ROSAT X-ray telescope PSPC and HRI detectors (Greiner et al., 1993), in the range 0.1 - 2.4 keV. Analysis of the ROSAT data shows that the hydrogen column density is on the order of $5 \times 10^{22} cm^2$ (Greiner et al., 1994). A variable radio source, consistent with the ROSAT position was found, whose light curve seems to be correlated with the hard X-ray time profile (Mirabel et al., 1993a). Recently, this object has been proven to be associated with superluminal radio jets (Mirabel and Rodriguez, 1994b). Two infrared sources were detected in the ROSAT HRI error box, one of them being consistent with the radio source (Mirabel et al., 1993b). It is interesting to note, that GRS1915+105 is still active, both at radio and hard X-ray wavelengths (Gérard et al., 1994; Sazonov et al., 1994; Nagase et al., 1994). In this respect, GRS 1915+105, with its 2 month intensity rise to the maximum of the light curve followed by strong variations, differs from most other X-ray transients.

X-ray sources displaying such hard variations deserve a singular interest, since they include all galactic black hole candidates (BHC) known so far (e.g. GX 339-4, A 0620-00, GS 2000+25, GS 1124-68, Cyg X-1). Some properties of GRS 1915+105 are reminiscent of GX 339-4, a well known BHC, which is a highly variable source. This object may be compared also with 1E 1740.7-2942, which is probably located in a giant molecular cloud complex and has a radio counterpart (Mirabel et al., 1992). In both cases there is some correlation between the radio and X-ray fluxes. Like GRS 1915+105, both 1E 1740.7-2942 and GX 339-4 have long periods of outburst. On the other hand, GX 339-4 had a very rapid rise time (10 days), followed by a nearly exponential decay.

This source may also be compared with binary systems containing a neutron star, which may have slow rise times and a similar spectral index. No QPO has been reported in the GRS 1915+105 light curve and, furthermore, if some analogy may be found with KS 1731-260 (Barret et al., 1992), no type I burst has been observed. An association with the Soft Gamma-ray Repeater SGR 1900+14 = B 1900+14 has been proposed (Grindlay, 1994; Mirabel et al., 1994a) on the basis of time and space proximity, though we note that the probability of such an association, when the solid angle and the duration of the GRS source are taken into account, is of the order of 20%. Clearly, GRS 1915+105 is a very peculiar source, whose nature is not clearly understood, and deserves special interest.

We present here spectroscopic and photometric optical observations of the objects located within the ROSAT HRI error box. Results of these observations are given in the next section, and we use these data to derive the nature of the objects found, whenever possible, and the hydrogen column density along the line of sight, which we then compare to the results obtained at X-ray, radio and infrared wavelengths.

## 2. The observations

The observations of GRS 1915+105 reported here were performed at the 3.6m Canada - France - Hawaii - Telescope (hereafter CFHT), at the ESO 3.5m New Technology Telescope (hereafter NTT) and at the Observatoire de Haute Provence 1.20m telescope (hereafter OHP). At CFHT the Cassegrain focus was equipped with the MOS/SIS imaging spectrograph. A Loral 2048x2048, 15$\mu$m pixel size i.e. 0.31 arcsec/pixel, CCD detector, and a 150Å/mm grism were used in this configuration. At the ESO, the EMMI spectrometer with a 246Å/mm grism was at NTT focus in conjunction with a Loral 2048x2048 CCD detector, for a total exposure time of 2 hours, on June 28, 1993. At the OHP, a Tektronix 512x512 CCD camera was used at the Newton focus of the 1.20 m telescope yielding a pixel size of 0.77 arcsec. A total of 1 hour data was acquired in the I band during four nights in the period June 25 - 29, 1993.

Photometric results presented here are based on the CFHT observations of August 8 and 9, 1993. OHP images were used to search for a possible long term variability in the I photometric band. The mean seeing was 1 arcsec for the first night, and 0.7 arcsec for the second night at CFHT, and 2.6 arcsec at OHP. Spectroscopy of objects 1, 3 and 4 was made at CFHT, while the spectrum of object 2 was obtained at the NTT. All spectra and images where reduced using the standard procedures of the MIDAS image processing system.

We show on figure 1 a corrected image of the ROSAT HRI error box in the I photometric band. Table 1 lists all detected objects, with their main photometric characteristics from CFHT data. Note that source 5 is at the position of the superluminal radio source and thus the likely optical counterpart. We could perform low resolution spectra of sources 1, 2 and 3 only, because of the faintness of the 2 remaining objects. These spectra are displayed on figures 2 to 4. For completeness we present the spectrum of source 4 on figure 5, while this object is too faint to allow any firm conclusions to be drawn. At OHP, objects 5 and 6 were not detected, and the magnitude of the sources 2 and 3 was at the same level than at CFHT. Because of the large seeing, objects 1 and 4 could not be separated.

## 3. Results

### 3.1 The spectra of the three brightest sources

*Object 1* : The presence of CaII, HeI, OI, and NaI lines suggest a B, or alternatively a G or K type star. Using the width of the 5780Å and 6284Å diffuse interstellar bands (Bromage and Nandy, 1973 ; Snell and Vanden Bout, 1981) we find that the B-V color excess of object 1 is 2.2 $\pm$ 0.12. If we use this excess to compute the intrinsic spectrum of object 1, then we can exclude a B type origin. Therefore we conclude that object 1 is probably a star of type F or G. The appearance of the Pashen lines seems to favour a spectral type earlier than G5 (Torres - Dodgen and Weaver, 1993).

*Object 2* : The absence of strong spectral features does not allow any firm conclusions. However, the presence of CaII and of Pashen lines favours a G spectral type.

*Object 3* : The presence of MgI, CaII, and NaI lines is characteristic of a late type spectrum, i.e. G to K3V or III. The visibility of molecular TiO bands seems to confirm this late origin. Again, using the 6284Å interstellar band we find $E_{B-V}$ of about 1.7.

Using the reddening of objects 1 and 3 we may compute the lower limits to the distances of these stars. Assuming a G5V type for star 1 and K2V for star 3, i.e. absolute magnitudes $M_V = 7$ and 6.4 respectively, we find a distance of at least 650pc (object 1) or 950pc (object 3).

### 3.2. The object 5 as the GRS 1915+105 source counterpart

The position of object 5 is consistent with that of the radio source. Mirabel et al. (1994) report that its K magnitude varies between 13 and 14, while we find $I = 23.4 \pm 0.4$. Using the new relation between the hydrogen column density and the optical absorption calibrated on well known LMXBs (Predehl and Schmitt, 1994), the reported $n_H$ of $5 \times 10^{22}$ (Greiner et al., 1994) corresponds to $A_V = 28$ magnitudes, consistent with the molecular emission line observations (F. Mirabel, private communication). This implies a reddening of $E_{B-V} = 9.6$, or equivalently, $E_{I-K} = 11.3$. Hence, the intrinsic color index of object 5 is $(I-K)_0 = -1.4 \pm 0.6$, assuming a mean K magnitude of 13.5, consistent with a hot object, and a massive companion cannot be excluded. From the absolute magnitude of the infrared counterpart ($M_K \approx -5.6$ for d = 12.5 kpc) we derive its I intrinsic magnitude, $M_I \approx -6.7$, and we note that our results are consistent with those obtained by Mirabel and Rodriguez (1994).

### 4. Discussion and conclusions

We have performed deep photometry and whenever possible spectroscopy of the objects found in the ROSAT HRI X-ray error box of GRS 1915+105. They are heavily absorbed and located at a distance of at least 1kpc.

The likely counterpart of the radio source has an intrinsic color index $(I-K)_0$ about -1.4. As a comparison, this index is 0.9 for the black hole candidate V404 Cyg (Casares et al., 1993). The absolute I magnitude of SS 433, $M_I \approx -7.5$, and color index, $(I-K)_0 \approx -0.2$ are quite close to that of GRS 1915+105, suggesting that both systems are the association of compact object with a hot, massive star.

Among the known hard X-ray source in the galactic bulge, 5 of them have been associated with radio jets, and only 3 display relativistic ejections of matter. The characteristics of SS 433 and GRS 1915+105 seem in that sense quite rare and similar, though some other systems of that type may be hidden by the interstellar extinction in the direction of the center of the Galaxy.

*Acknowledgements* :

We thank Dr. François Hammer (CFHT / CNRS) for his help in using the MOS/SIS instrument at CFHT. Many thanks are also due to Dr J. Storm who took the frames at ESO/NTT and to Dr. K. Reinsch who acquired the images at the Observatoire de Haute Provence (CNRS). Observing time at ESO/NTT was awarded on director discretionary time.

# References

Awaki H, et al., 1991, PASJ 43, 195

Barret D., et al., 1992, ApJ 394, 615

Bromage G.E., and Nandy K., 1973, A&A, 26,17

Casares, J., et al., 1993, MNRAS, 265, 834

Castro-Tirado A.J., et al., 1992, IAUC 5590

Castro-Tirado A.J., et al., 1994, ApJS 92, 469

Gérard E., Rodriguez L.F., Mirabel I.F., 1994, IAUC 5958

Greiner J., et al., 1993, IAUC 5786

Greiner J., et al., 1994, Proceedings of the Second Compton Symposium, College Park, 1993, eds. C.E. Fichtel, N. Gehrels, and J.P. Norris, AIP Conference Proceedings 304, AIP Press, New York, p. 260

Grindlay, J.E., 1994, ApJS, 92, 465

Greiner J., et al., 1994, Proceedings of the Second Compton Symposium, College Park, 1993, eds. C.E. Fichtel, N. Gehrels, and J.P. Norris, AIP Conference Proceedings 304, AIP Press, New York, p. 210

Mirabel I.F., et al., 1992, Nat 358, 215

Mirabel I.F., et al., 1993a, IAUC 5773

Mirabel I.F., et al., 1993b, IAUC 5830

Mirabel I.F., Duc P.A., Rodriguez L.F., Teyssier R., Paul J., Claret A., Auriere M., Golombek D., Marti J., 1994, A&A 282, L17

Mirabel I.F., and Rodriguez, L.F., 1994, Nat 371, 46

Nagase F., et al., 1994, IAUC 6094

Predehl P., and Schmitt J.H.M.M., 1995, A&A, in press

Sazonov S., Sunyaev R., Lapshov I., 1994, IAUC 5959

Snell R.L., and Vanden Bout P.A., 1981, ApJ, 244, 844

Torres - Dodgen A.V., and Weaver Wm.B., 1993, PASP, 105, 693

**Figure Captions**

Figure 1 : I band image of the ROSAT HRI error box for GRS 1915+105

Figure 2 : spectrum of object 1.

Figure 3 : spectrum of object 2.

Figure 4 : spectrum of object 3.

Figure 5 : spectrum of object 4.

| Object | Apparent magnitude | | | |
|---|---|---|---|---|
| | B | V | R | I |
| 1 | $> 25.9$ | $22.92 \pm 0.19$ | $20.79 \pm 0.17$ | $17.25 \pm 0.15$ |
| 2 | $> 25.9$ | $22.30 \pm 0.18$ | $20.93 \pm 0.17$ | $17.54 \pm 0.15$ |
| 3 | $> 25.9$ | $21.65 \pm 0.17$ | $20.35 \pm 0.16$ | $17.80 \pm 0.15$ |
| 4 | $> 25.9$ | $> 26.1$ | $25.8 \pm 0.3$ | $21.23 \pm 0.18$ |
| 5 | $> 25.9$ | $> 26.1$ | $> 26.1$ | $23.42 \pm 0.30$ |
| 6 | $> 25.9$ | $> 26.1$ | $> 26.1$ | $24.9 \pm 0.5$ |

Table 1 : Photometric characteristics of sources found in the HRI error box